\documentclass[twocolumn, prb, aps, 10pt, superscriptaddress]{revtex4-2}

\usepackage{graphicx}
\usepackage{amsmath}
\usepackage{amsfonts}
\usepackage{amssymb}
\usepackage{mathrsfs}
\usepackage{hyperref}
\hypersetup{colorlinks = true, allcolors = blue}
\usepackage{xr}
\usepackage[T1]{fontenc}
\usepackage[utf8]{inputenc}

\newcommand{\nn}{\nonumber\\}

\newcommand{\PP}[1]{\Phi_{#1}}
\newcommand{\PG}[1]{\Phi_{#1}^{\mathrm{G}}}
\newcommand{\PX}[1]{\Phi_{#1}^{\mathrm{X}}}
\newcommand{\Pp}[1]{\Phi_{#1}^{+}}
\newcommand{\Pm}[1]{\Phi_{#1}^{-}}
\newcommand{\Ppm}[1]{\Phi_{#1}^{\pm}}

\newcommand{\wL}{\omega_{\mathrm{L}}}
\newcommand{\wC}{\omega_{\mathrm{C}}}

\newcommand{\maintitle}{Generation and manipulation of photon number wave packets in photonic cavities}

\begin{document}

\title{\maintitle}
\author{L.~Nimmesgern}
\affiliation{Lehrstuhl f\"{u}r Theoretische Physik III, Universit{\"a}t Bayreuth, 95440 Bayreuth, Germany}
\author{M.~Cygorek}
\affiliation{Condensed Matter Theory, Department of Physics, TU Dortmund, 44221 Dortmund, Germany}
\author{D.~E.~Reiter}
\affiliation{Condensed Matter Theory, Department of Physics, TU Dortmund, 44221 Dortmund, Germany}
\author{V.~M.~Axt}
\affiliation{Lehrstuhl f\"{u}r Theoretische Physik III, Universit{\"a}t Bayreuth, 95440 Bayreuth, Germany}

\begin{abstract}
Quantum emitters inside optical cavities can create not only fixed photon number states but also photon number wave packets, which are states with a finite photon number distribution that oscillates in time.
These states emerge when the emitter is driven by an external field while coupled to the cavity.
We show that by rapidly changing the driving strength, new wave packets can be generated, allowing multiple packets to coexist and evolve independently.
We classify the resulting wave packet behavior into distinct dynamical subclasses between which we choose through the choice of relevant parameters.
Based on this understanding, we develop simple and robust protocols to generate a specified number of photon number wave packets on demand.
We propose that the rich dynamics can be experimentally investigated by merely measuring the mean photon number.
\end{abstract}

\maketitle

\section{Introduction}
\label{sec:introduction}
The infinitely many discrete Fock-number-states of a single photon mode in a cavity can be regarded as a large register for storing information that may be further used for quantum information processing~\cite{ralph_quantum_2003, zoller_quantum_2005, acin_quantum_2018, vajner_quantum_2022}.
The information can be encoded in any structured excitation of the number states.
In order to be useful, different structures must be easy to distinguish and it should be possible to read out the information.
The latter can be accomplished by using advanced experimental techniques that allow for direct recording of the complete number state of photon modes~\cite{schlottmann_exploring_2018, schmidt_photon-number-resolving_2018, helversen_quantum_2019}.

\begin{figure}[t]
	\centering
	\includegraphics[scale=1]{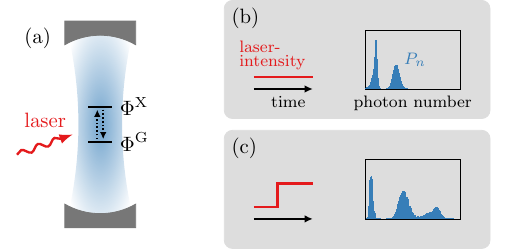}
	\caption{
		(a) Sketch of the system studied in this work.
		A two-level system (TLS) coupled to a single mode of a photonic cavity.
		Transitions between the ground state $\PG{}$ and the excited state $\PX{}$ of the TLS are induced additionally by an external laser.
		(b), (c) Snapshots in time of the photon number distribution $P_{n}$ for constant driving intensity (b) and after a rapid increase in the intensity (c).
		In both cases the external laser is slightly detuned with respect to the cavity frequency.
	}
	\label{fig:introduction}
\end{figure}

The idea of encoding information in the photon number distribution is far from being mature in particular because efficient and easy to use techniques to generate structured excitations in the number state distribution are currently not well developed.
In principle, it is possible to prepare an arbitrary state of a photon mode in a cavity by coupling the mode to a driven two-level system (TLS)~\cite{law_arbitrary_1996} as sketched in Fig.~\ref{fig:introduction}(a).
A TLS can be realized in a multitude of physical systems, e.g. semiconductor quantum dots~\cite{reithmaier_strong_2008, kasprzak_coherence_2013, bozzio_enhancing_2022, karli_super_2022, spinnler_single-photon_2024}, superconducting qubits~\cite{hofheinz_generation_2008, hofheinz_synthesizing_2009, arute_quantum_2019} or atomic systems~\cite{shore_jaynes-cummings_1993, blinov_quantum_2004, haffner_quantum_2008, pogorelov_compact_2021}.
However, the protocol proposed in Ref.~\cite{law_arbitrary_1996} requires both the driving of the TLS as well as its coupling to the photon mode to be time-dependent and controllable which makes this proposal difficult to implement.
Furthermore, preparing nontrivial structures typically requires many steps.
This makes the protocol susceptible to disturbances which can even cause a complete failure of the preparation scheme~\cite{cosacchi_schrodinger_2021}.
One possible strategy to circumvent these problems is to consider only special structures in the photon number distributions which are easy to prepare and which persist for an extended period of time after preparation.
These demands are fulfilled by photon number wave packets.
A photon number wave packet is a structure in the photon number distribution centered around some finite mean photon number and extending only over a narrow region of neighboring photon numbers.
Such a wave packet can evolve in time running up and down the ladder of photon number states.

It has long been noted that a single photon number wave packet can be generated simply by driving the TLS by a strong cw field in resonance with the cavity frequency~\cite{chough_nonlinear_1996}.
Also, using chirped pulses for the excitation, wave packet-like structures can appear in the photon number distribution exhibiting two maxima~\cite{cosacchi_transiently_2020}.
However, the underlying mechanism is restricted to the generation of at most two coexisting packets.
As recently discovered, by driving the TLS with a strong but slightly off-resonant cw field [cf. Fig.~\ref{fig:introduction}(b)] two or more photon number wave packets can be prepared simultaneously~\cite{nimmesgern_multiple_2024}.

In the present paper we demonstrate that by suddenly switching the strength of the driving it is possible to produce on demand new photon wave packets thereby increasing the number of coexisting packets at will [cf. Fig.~\ref{fig:introduction}(c)].
Apart from its potential to be useful for information processing, it turns out that the process of packet generation is by itself an interesting physical phenomenon.
In fact, we find a rich variety of dynamical scenarios where depending on the chosen parameters a sudden switch may or may not lead to an additional wave packet.
By analyzing the pertinent conditions that determine the dynamical behavior we derive criteria under which circumstances specific dynamical regimes are found.
Our work provides a new toolbox for targeted engineering of the photon number distribution where at prescribed moments in time new structures can be induced.

After briefly introducing the model in Sec.~\ref{sec:model}, we analyse the photon number wave packet dynamics for constant driving in Sec.~\ref{sec:constant_f}.
Extending previous results~\cite{nimmesgern_multiple_2024}, we apply a variational approach and classify the occuring dynamics based on detuning and driving strength.
In Sec.~\ref{sec:control}, we then propose control protocols based on a single or multiple rapid steps for on-demand photon number wave packet generation.
We then show in Sec.~\ref{sec:measurement} how to experimentally extract the packet structure before concluding in Sec.~\ref{sec:summary}.

\section{Theoretical model}
\label{sec:model}
The TLS-cavity system is described by a Jaynes--Cummings model \cite{gerry_introductory_2004}.
The ground and excited states of the TLS are denoted $\PG{}$ and $\PX{}$, respectively, while $\PP{n}$ are the photonic Fock states.
We write the corresponding product states as $\PG{n} = \PG{} \otimes \PP{n}$ and $\PX{n} = \PX{} \otimes \PP{n}$.
The Hamiltonian, expressed in the rotating frame with respect to the driving frequency $\wL$, takes the form
\begin{align}\label{eqn:hamiltonian}
	H
	={}& \hbar \delta\, a^{\dagger} a
	+ \hbar g\, \bigl(a \sigma_{+} + a^{\dagger} \sigma_{-}\bigr)\nn
	{}&- \hbar f(t)\, \bigl(\sigma_{+} + \sigma_{-}\bigr).
\end{align}
Here, $a$ ($a^{\dagger}$) is the photonic annihilation (creation) operator and $\sigma_{+}$, $\sigma_{-}$ are the ladder operators of the TLS.
The TLS and external driving are assumed to be in resonance, while $\delta = \wC - \wL$ is the difference between the frequency of the cavity mode $\wC$ and $\wL$.
The coupling strength between TLS and cavity is denoted by $g$ and the time-dependent driving strength by $f(t)$.
In this work, we focus on the strong driving regime, i.e. we choose parameters that fulfill the hierarchy $\delta \ll g \ll f$.

The wave function $\Psi(t)$ is obtained by expanding the Schr\"{o}dinger equation
\begin{align}\label{eqn:schrodinger}
	i \hbar\, \dot{\Psi} = H \Psi
\end{align}
in the bare state basis $\{\PG{n}, \PX{n}\}$ and solving the resulting set of coupled equations truncated at a suitably chosen highest photon number with a classic Runge--Kutta method (RK4).
If not stated otherwise, the initial state is chosen as $\Psi(0) = \PG{0}$, i.e. the ground state of the TLS with zero photons in the cavity.

\section{Analysis of constant driving}
\label{sec:constant_f}
In this section, we discuss the dynamical behavior of the system for constant driving strength $f$.
The solutions for constant driving strength are discussed in detail in Ref.~\cite{nimmesgern_multiple_2024} focusing on qualitative understanding through the use of an effective Hamiltonian and a discrete WKB method.
After briefly summarizing the results of this analysis, we develop a novel approximation of the same situation based on a variational principle resulting in a quantitative description of the packet dynamics.
Understanding the case $f = \mathrm{const.}$ is a key step towards understanding how to generate new packets in a controlled manner.

\begin{figure}[t]
	\centering
	\includegraphics[scale=1]{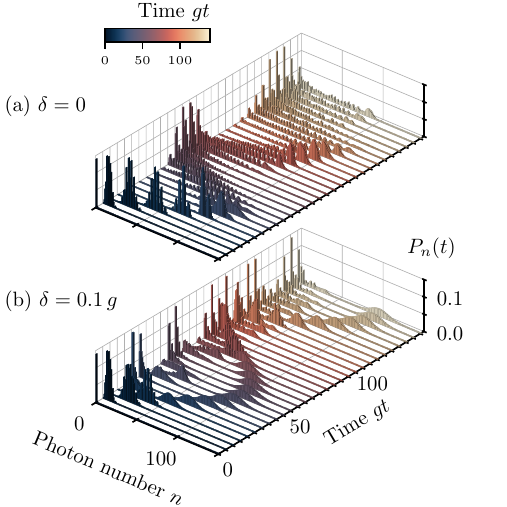}
	\caption{
		Time evolution of the photon number distribution $P_{n}(t)$ visualized by plotting $P_{n}(t)$ at constant $t$ for several snapshots in time.
		The solutions were obtained by numerically integrating Eq.~\eqref{eqn:schrodinger} with a constant driving strength of $f = 5\, g$ and a detuning of $\delta = 0$ (a) and $\delta = 0.1\, g$ (b).
	}
	\label{fig:slices}
\end{figure}

\subsection{Summary of previous findings}
\label{ssec:constant_f:summary}
In Ref.~\cite{nimmesgern_multiple_2024}, it was found that the wave function consists of two parts
\begin{align}
	\Psi(t) = \frac{1}{\sqrt{2}}\Bigl(\Psi_{1}(t) + \Psi_{2}(t)\Bigr),
\end{align}
where $\Psi_{1}$ ($\Psi_{2}$) are predominantly composed of eigenstates of $H$ with eigenenergies centered around $-\hbar f$ ($\hbar f$).
$\Psi_{1}$ and $\Psi_{2}$ manifest themselves as distinct packets in the photon number distribution $P_{n}(t) = |\langle \PG{n} | \Psi(t) \rangle|^{2} + |\langle \PX{n} | \Psi(t) \rangle|^{2}$.
Constant driving induces the mean of these structures of finite width in $P_{n}$ to oscillate.
At zero detuning $\delta = 0$, the oscillation amplitude as well as the frequency of both packets are equal.
Hence, they cannot be distinguished in the photon number distribution [cf. Fig.~\ref{fig:slices}(a)].
Fig.~\ref{fig:slices}(b) shows the evolution of $P_{n}$ for nonvanishing detuning.
Here, the amplitudes and frequencies can differ strongly depending on the precise choice of parameters and thus $P_{n}$ exhibits two separately evolving maxima.
On top of the oscillation, the packets disperse during their evolution leading to an eventual breakdown of the packet as a well-defined concept in the photon number distribution.
It is worthwhile to note that this dispersion is much weaker at finite detuning, thus making packets generated at finite $\delta$ more robust.
Moreover, it was found that for $\delta \approx g^{2}/f$, $\Psi_{2}$ splits into multiple individual packets during the course of its evolution, resulting in a continually rising number of packets.

As a result of strong driving, $f \gg g$, for low photon numbers $n \ll (4f/g)^{2}$ to a good approximation each packet is a product state of some photonic state and a single laser-dressed state (LDS)
\begin{align}\label{eqn:lds_def}
	\Ppm{} = \frac{1}{\sqrt{2}}\, \bigl(\PG{} \pm \PX{}\bigr).
\end{align}
If the detuning is sufficiently large, the oscillation amplitude is low enough such that this condition holds throughout the evolution.
In this LDS-decoupled regime, the photon number distribution of a single packet resembles that of a Poissonian.
However, in general the wave packets can deform dramatically resulting in a highly nonclassical state~\cite{nimmesgern_multiple_2024}.

\begin{figure*}[t]
	\centering
	\includegraphics[scale=1]{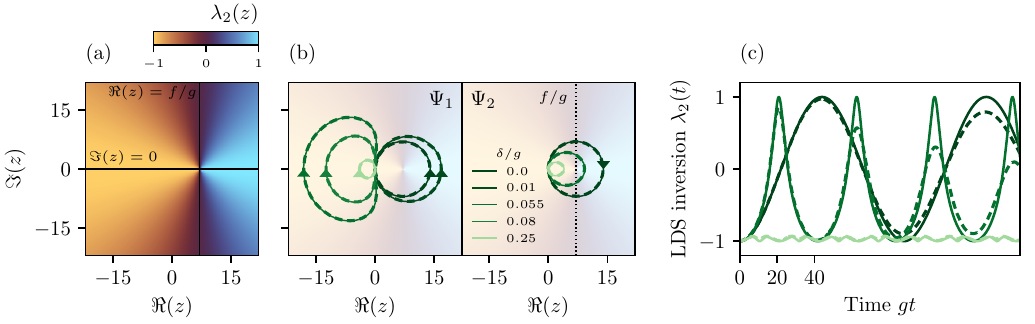}
	\caption{
		(a) Overlap between the variational eigenvectors and the TLS $\lambda_{2}$ as a function of $z$ calculated via Eq.~\eqref{eqn:variational_lds}.
		(b) Several trajectories of $z_{1}$ (left panel) and $z_{2}$ (right panel) obtained by numerically integrating Eq.~\eqref{eqn:variational_eom_z_adiab} using a driving strength of $f = 7\, g$ and values of the detuning $\delta / g$ of $0$, $0.01$, $0.055$, $0.08$ and $0.25$ (solid lines).
		For the sake of visual clarity only the solutions for $\delta/g = 0$, $0.055$ and $0.25$ are shown for $z_{2}$.
		In comparison, the trajectories of the maximum of the absolute value of the Wigner function $W(z)$ during the first oscillation cycles are shown (dashed lines).
		They were obtained by numerically integrating Eq.~\eqref{eqn:schrodinger} with the same parameters and initial state $\Psi(0) = \Pp{0}$ (left panel) and $\Psi(0) = \Pm{0}$ (right panel).
		(c) $\lambda_{2}$ as a function of time $t$ along trajectories of $z_{2}$ for the same parameters as in (b).
		$z_{2}$ was obtained by numerically integrating Eq.~\eqref{eqn:variational_eom_z_adiab} and $\lambda_{2}$ by subsequent application of Eq.~\eqref{eqn:variational_lds}
		The dashed lines show the LDS inversion calculated from numerical solutions of Eq.~\eqref{eqn:schrodinger} using the same parameters and the initial state $\Psi(0) = \Pm{0}$.
	}
	\label{fig:variational_theory}
\end{figure*}

\subsection{Variational approximation of a single packet}
The simplifying methods used in Ref.~\cite{nimmesgern_multiple_2024} are each restricted to respective ranges of photon numbers or choices of parameters.
They are thus useful to gain qualitative understanding, but not for extensive quantitative analysis.
In this work, we are interested in quantitative descriptions of the packet mean and the TLS state, since this allows us to predict under which circumstances we can generate an additional packet.
To this end, we develop a new approximate description based on a variational principle.
The crudest approximation, that is still able to describe the packet structure observed in numerical calculations, consists in assuming each single packet to be a product state of an arbitrary TLS state, described by $\alpha$ and $\beta$, and a coherent state with the coherent amplitude $z$:
\begin{align}\label{eqn:variational_ansatz}
	\Psi_{j}
	= \Bigl(\alpha_{j} \PG{} + \beta_{j} \PX{}\Bigr)
	\otimes \exp\Bigl(-\frac{|z_{j}|^{2}}{2}\Bigr) \sum_{n=0}^{\infty} \frac{z_{j}^{n}}{\sqrt{n!}} \PP{n}.
\end{align}
This Ansatz neglects packet deformation, as well as any entanglement between the TLS and the photonic system.
While the former is generally a rough approximation and the latter is only strictly justified in the LDS-decoupled regime, the variational approximation using this Ansatz nevertheless describes the main features satisfactorily, as shown below.

The core idea of the variational approximation is to determine the trajectory $\alpha(t), \beta(t), z(t)$ such that Eq.~\eqref{eqn:variational_ansatz} approximates solutions of the Schr\"{o}dinger equation optimally.
This procedure results in the equations of motion (cf. Appendix~\ref{app:variational})
\begin{subequations}\label{eqn:variational_eom}
\begin{align}
	\label{eqn:variational_eom_tls}
	i \hbar \begin{pmatrix}
		\dot{\alpha}_{j}\\
		\dot{\beta}_{j}
	\end{pmatrix} &= H_{\mathrm{TLS}}(z_{j}) \begin{pmatrix}
		\alpha_{j}\\
		\beta_{j}
	\end{pmatrix}
	\\
	\label{eqn:variational_eom_z}
	i \hbar\, \dot{z}_{j}
	&= \hbar \delta\, z_{j} + \hbar g\, \alpha_{j}^{*} \beta_{j},
\end{align}
\end{subequations}
where
\begin{align}\label{eqn:variational_HTLS}
	H_{\mathrm{TLS}}(z)
	= \begin{pmatrix}
		0 & \hbar g\, z^{*} - \hbar f\\
		\hbar g\, z - \hbar f & 0
	\end{pmatrix}.
\end{align}
For sufficiently slowly varying $z(t)$, the solution for the TLS state is given by the adiabatic approximation, which we describe in the following.

If $z_{j}$ were constant, the solutions of Eq.~\eqref{eqn:variational_eom_tls} would be given by the eigenvectors $\varphi_{1}$ and $\varphi_{2}$ of $H_{\mathrm{TLS}}$ and their respective eigenfrequencies
\begin{align}\label{eqn:variational_eigenfrequency}
	\omega_{1/2}(z) = \mp |g z - f|.
\end{align}
For purely real values of $z$ these eigenstates coincide with the LDS.
A nonzero imaginary part in $z$ changes the relative phase of $\varphi_{1/2}^{\mathrm{G}}$ and $\varphi_{1/2}^{\mathrm{X}}$, the ground and excited state amplitudes of $\varphi_{1/2}$, while they still fulfill $|\varphi_{1/2}^{\mathrm{G}}| = |\varphi_{1/2}^{\mathrm{X}}|$.
We characterize the overlap between the variational eigenstates and the LDS by
\begin{align}
	\lambda_{1/2}(z)
	&= |\langle \Pp{} | \varphi_{1/2} \rangle|^{2} - |\langle \Pm{} | \varphi_{1/2} \rangle|^{2}.
\end{align}
For a given coherent amplitude $z$, this value is given by (cf. Appendix~\ref{app:variational})
\begin{align}\label{eqn:variational_lds}
	\lambda_{1/2}(z)
	= \mp \mathrm{sgn}(\Re(z) - f/g) \Bigl(1 + \frac{\Im(z)^{2}}{(\Re(z) - f/g)^{2}}\Bigr)^{-1/2}.
\end{align}
Note that due to the normalization condition
\begin{align}
	|\varphi_{1/2}^{\mathrm{G}}|^{2} + |\varphi_{1/2}^{\mathrm{X}}|^{2} = 1,
\end{align}
the individual LDS occupations can be directly calculated from $\lambda$.
When $z_{j}(t)$ is slowly varying, the solutions of Eq.~\eqref{eqn:variational_eom_tls} follow these eigenvectors adiabatically~\cite{born_beweis_1928}.
Then we find
\begin{align}\label{eqn:variational_adiabatic}
	\begin{pmatrix}
		\alpha_{j}(t)\\
		\beta_{j}(t)
	\end{pmatrix}
	&= \varphi_{j}(z_{j}(t))\, \exp\Bigl(-i \int_{0}^{t} \omega_{j}(z_{j}(t'))\, \mathrm{d}t'\Bigr),
\end{align}
with $j \in \{1, 2\}$, to be good approximate solutions.
Inserting these expressions into Eq.~\eqref{eqn:variational_eom_z}, the equation of motion for $z$ takes the form
\begin{align}\label{eqn:variational_eom_z_adiab}
	i \hbar\, \dot{z}_{1/2}
	&= \hbar \delta\, z_{1/2} \mp \frac{\hbar g}{2} \frac{z_{1/2} - f/g}{|z_{1/2} - f/g|}.
\end{align}

Both the time-dependent variational approximation as well as the subsequent adiabatic approximation do not violate conservation of energy.
Hence, the average value of $H$
\begin{align}\label{eqn:variational_Hconst}
	\langle H \rangle_{1/2} = \hbar \delta\, |z_{1/2}|^{2} \mp \hbar\, |g z_{1/2} - f|
\end{align}
using the Ansatz Eq.~\eqref{eqn:variational_ansatz} and the adiabatic solution [cf. Eq.~\eqref{eqn:app:eigenvector_phase} in Appendix~\ref{app:variational}] is a constant of motion.
With the initial condition $z(0) = 0$, this yields
\begin{align}\label{eqn:variational_Hvalue}
	\langle H \rangle_{1/2} = \mp \hbar f.
\end{align}
Thus, the solutions $(\alpha_{1}, \beta_{1}, z_{1})$ and $(\alpha_{2}, \beta_{2}, z_{2})$ can be identified with the packets $\Psi_{1}$ and $\Psi_{2}$ respectively (cf. Sec.~\ref{ssec:constant_f:summary}) and they describe their behavior well insofar the deformation of the packets is neglected.

\subsection{TLS state within the variational approximation}
Along the adiabatic solutions, Eq.~\eqref{eqn:variational_lds} gives a direct relation between the photonic and TLS states, which is illustrated in Fig.~\ref{fig:variational_theory}(a).
If the numerical factor on the right side of this expression is equal or close to $1$, each eigenvector $\varphi_{1/2}$ essentially corresponds to one LDS.
This occurs when $\Im(z) \ll \Re(z) - f/g$, which is fulfilled in particular if $|z| \ll f/g$ or $\Im(z) = 0$.
The sign determines which eigenvector corresponds to which LDS.
If $\Re(z) < f/g$, $\varphi_{1}$ ($\varphi_{2}$) is predominantly constructed of $\Phi^{+}$ ($\Phi^{-}$), whereas the roles are reversed if $\Re(z) > f/g$.
On the other hand, if the numerical factor is significantly smaller than $1$, both eigenvectors are constructed of both LDS to considerable amounts.
Most notably, the LDS occupations are precisely equal whenever $\Re(z) = f/g$.
Using these observations, we can obtain a clear picture of the dynamical evolution of the TLS state after understanding the evolution of the photonic state.

\begin{figure*}[t]
	\centering
	\includegraphics[scale=1]{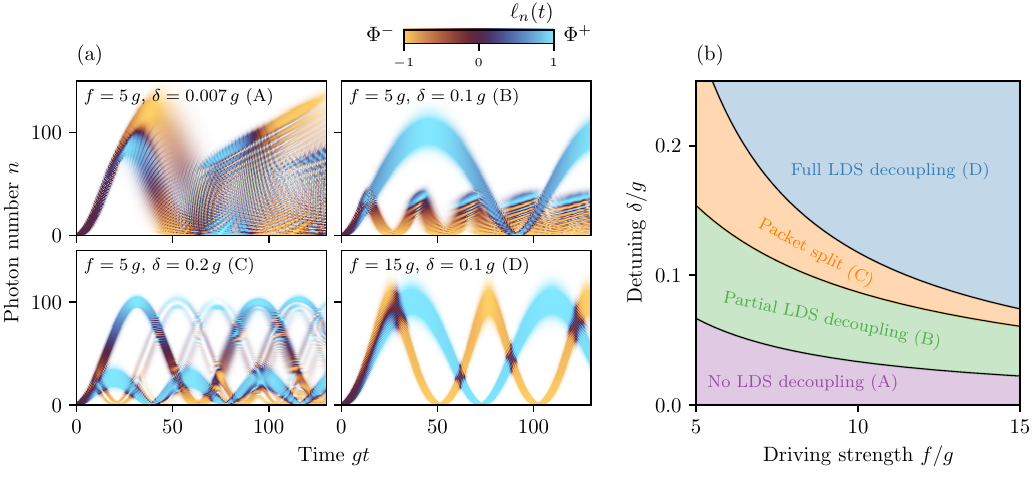}
	\caption{
		(a) Numerical solutions of Eq.~\eqref{eqn:schrodinger} using parameters (A) $f = 5\, g$, $\delta = 0.007\, g$, (B) $f = 5\, g$, $\delta = 0.1\, g$, (C) $f = 5\, g$, $\delta = 0.2\, g$ and (D) $f = 15\, g$, $\delta = 0.1\, g$.
		Each set of parameters serves as an illustration of the behavior exhibited by the system in the respective classes A, B, C, D.
		The measure $\ell_{n}(t)$ [cf. Eq.~\eqref{eqn:lds_measure}] resolved in photon number $n$ and time $t$ is shown in a linear color scale, while the opacity is given by the photon number occupation $P_{n}(t)$.
		(b) Sketch of the parameter ranges for the individual classes.
		The boundary between classes A and B is drawn as $f \delta = g^{2} / 3$ and those between B, C and C, D as $(f + 1.5\, g) \delta = g^{2}$ and $(f - 1.5\, g) \delta = g^{2}$ respectively.
	}
	\label{fig:constant_f}
\end{figure*}

\subsection{Photonic state within the variational approximation}
Eq.~\eqref{eqn:variational_eom_z_adiab} is a single equation from which the photonic trajectory $z(t)$ can be calculated numerically.
Fig.~\ref{fig:variational_theory}(b) shows its solutions for several values of the detuning.
We recall that for a coherent state, as assumed in Eq.~\eqref{eqn:variational_ansatz}, the coherent amplitude $z$ marks the position of the maximum of the Wigner function $W$ in phase space, where $W$ is defined as~\cite{barnett_methods_2002}
\begin{align}
	W(z)
	= \frac{2}{\pi} \mathrm{Tr}\Bigl((-1)^{a^{\dagger} a} D^{\dagger}(z) \rho_{\mathrm{phot}} D(z)\Bigr).
\end{align}
Thus, in order to check the validity of our approximation we compare $z(t)$ found from Eq.~\eqref{eqn:variational_eom_z_adiab} with the trajectory of the maximum of $|W(z)|$ of the corresponding packet by numerically integrating Eq.~\eqref{eqn:schrodinger}.
As seen in Fig.~\ref{fig:variational_theory}(b), the agreement is almost perfect for the first oscillation cycle although deviations from the Poisson shape of the distribution are obvious from Fig.~\ref{fig:slices}.

In general, $z_{1/2}(t)$ describes oscillations between the initial state $z_{1/2}(0) = 0$ and a real turning point $\tilde{z}_{1/2}$ which corresponds to the maximal value of the photon number $|z|^{2}$.
These are derived in Appendix~\ref{app:turning}.
In the resonant case $z_{1}$ and $z_{2}$ share the common turning point $\tilde{z}_{1/2} = 2f/g$.
If $\delta$ is sufficiently small but nonzero, $\tilde{z}_{1}$ increases and $\tilde{z}_2$ decreases while still remaining greater than $f/g$.
According to Eq.~\eqref{eqn:variational_lds}, the TLS state of packet $\Psi_{1}$ ($\Psi_{2}$) therefore oscillates between $\Pp{}$ ($\Pm{}$) at $z = 0$ and $\Pm{}$ ($\Pp{}$) at the upper turning point.
At $\delta = g^{2}/8f$, $\tilde{z}_{1}$ discontinuously changes to negative values [cf. Eq.~\eqref{eqn:app:turning_1}].
Accordingly, the real part of $z_{1}$ is negative for considerable amounts of its trajectory.
In fact, $\Re(z_{1})$ is non-positive for all $t$ if $\delta > g^{2}/2f$.
Since the imaginary part of $z_{1}$ is not significantly larger than its real part, the single LDS $\Pp{}$ dominantly contributes to the TLS state of $\Psi_{1}$ throughout its evolution according to Eq.\eqref{eqn:variational_lds}.
On the other hand, $\tilde{z}_{2}$ decreases continuously as the detuning increases, crossing $\tilde{z}_{2} = f/g$ at $\delta = g^{2}/f$ [cf. Eq.~\eqref{eqn:app:turning_2}].
When the turning point is positive but sufficiently smaller than $f/g$, Eq.~\eqref{eqn:variational_lds} similarly predicts $\Psi_{2}$ to be constructed predominantly from the single LDS $\Pm{}$.
The dynamical evolution of $\lambda$ as described by Eqs.~\eqref{eqn:variational_lds} and \eqref{eqn:variational_eom_z_adiab} is compared to the LDS inversion $\sum_{n=0}^{\infty} (|\langle \Pp{n} | \Psi \rangle|^{2} - |\langle \Pm{n} | \Psi \rangle|^{2})$ obtained via numerical solutions of Eq.~\eqref{eqn:schrodinger} in Fig.~\ref{fig:variational_theory}(c).
We see satisfactory agreement with exception to the fact that the exact solutions do not exhibit full inversion at the maxima of photonic oscillations.
In addition, the agreement worsens during the evolution which can be traced back to the continued packet dispersion.

It is important to note that the adiabatic approximation, Eq.~\eqref{eqn:variational_adiabatic}, breaks down in the neighborhood of $z = f/g$ since here the difference of the eigenfrequencies of $\varphi_{1}$ and $\varphi_{2}$ vanishes [cf. Eq.~\eqref{eqn:variational_eigenfrequency}].
As a result, the LDS state is a superposition of $\varphi_{1}$ and $\varphi_{2}$ after passing through this region.
Within the confines of the Ansatz Eq.\eqref{eqn:variational_ansatz}, this superposition is forced to evolve as a single packet.
However, the actual solution has the parts corresponding to $\varphi_{1}$ and $\varphi_{2}$ evolve independently from each other according to Eq.~\eqref{eqn:variational_eom_z_adiab} respectively.
Thus, the packet splits after passing through $z \approx f/g$ and continues to do so, as the trajectory traverses this point again after a full oscillation.
For $z_{2}$ this situation sets in if $\delta \approx g^{2}/f$.

\subsection{Classification of the dynamical behavior}
The preceding analysis paints the picture of four distinct dynamical classes by which we can categorize the systems qualitative behavior.
Each class corresponds to a regime in parameter space spanned by driving strength and detuning, as sketched in Fig.~\ref{fig:constant_f}(b).
We denote the classes A, B, C and D, reflecting the order in which they occur with increasing $\delta$.
We emphasize that the transitions between classes are not well-defined at singular sets of parameters, but rather occur in continuous fashion.

In class A, at or close to $\delta = 0$, both packets experience significant contribution of both LDS in the course of their evolution.
Increasing $\delta$ at constant $f$ leads to first $\Psi_{1}$ and then $\Psi_{2}$ being decoupled into subspaces spanned by a single LDS.
The situation of a single LDS-decoupled packet is called class B.
In the full LDS-decoupled case we speak of class D.
Class C is characterized by a split of $\Psi_{2}$ into multiple packets and is thus located in parameter space around $f \delta = g^{2}$.
Representations of exemplary solutions for each class are shown in Fig.~\ref{fig:constant_f}(a) using the measure for LDS occupation
\begin{align}\label{eqn:lds_measure}
	\ell_{n}(t)
	= \frac{|\langle \Pp{n} | \Psi(t) \rangle|^{2} - |\langle \Pm{n} | \Psi(t) \rangle|^{2}}{
	|\langle \Pp{n} | \Psi(t) \rangle|^{2} + |\langle \Pm{n} | \Psi(t) \rangle|^{2}}.
\end{align}
The boundary between classes A and B can be deduced by inspecting $\lambda_{1}$.
For the sake of definiteness, we choose any parameter set, for which $\lambda_{1} > \frac{1}{2}$ throughout the trajectory $z_{1}(t)$, to belong to class B.
Then, the boundary between A and B is given by $f \delta = g^{2} / 3$ (cf. Appendix~\ref{app:classes}).
Similarly, we choose any solution with $\lambda_{2} < -\frac{1}{2}$ on $z_{2}(t)$ to belong to class D and therefore locate the transition between B and D at $f \delta = g^{2}$ (cf. Appendix~\ref{app:classes}).
Thus, class C represents the boundary between B and D.
By inspecting numerical solutions of Eq.~\eqref{eqn:schrodinger}, we establish the former to lie between $(f + 1.5\, g) \delta = g^{2}$ and $(f - 1.5\, g) \delta = g^{2}$.

\section{Control via time-dependent driving}
\label{sec:control}
Using our understanding of the case $f = \mathrm{const.}$, we now investigate mechanisms to control the number of photon number wave packets.
For this goal, we modulate the driving strength in a number of finite instantaneous steps.

\subsection{Single step}
\label{ssec:single}
Our first aim is to induce a third packet with a single step in the driving strength.
We thus take
\begin{align}
	f(t) = \begin{cases}
		f_{0}\, &\text{if } t < \tau
		\\
		f_{1}, &\text{if } t \geq \tau
	\end{cases}.
\end{align}
Since the Hamiltonian is time-independent for all $t \neq \tau$, the wave function for $t > \tau$ is equal to the solution of the Schr\"{o}dinger equation with constant driving strength $f_{1}$ and the initial state $\Psi(\tau)$.
Similarly, $\Psi(\tau)$ is the solution of the Schr\"{o}dinger equation with constant driving strength $f_{0}$ and an initial state of $\Psi(0) = \PG{0}$.

The states $\varphi_{1}$ and $\varphi_{2}$ in general depend on the driving strength.
Hence, $\Psi_{1}(\tau)$ and $\Psi_{2}(\tau)$ contain contributions of both modes $\varphi_{1}(f_{1})$ and $\varphi_{2}(f_{1})$.
Thus, within the realm of the variational approximation, the wave function for $t > \tau$ can be interpreted as consisting of four distinct packets where their relative weights are determined by the scalar products $\langle \varphi_{j}(f_{0}), \varphi_{k}(f_{1})\rangle$ evaluated at $z_{j}(\tau)$.
We denote by $z_{j,k}$ the trajectory of the packet which corresponds to eigenvector $\varphi_{j}$ for $t < \tau$ and $\varphi_{k}$ for $t > \tau$.

\begin{figure}[t]
	\centering
	\includegraphics[scale=1]{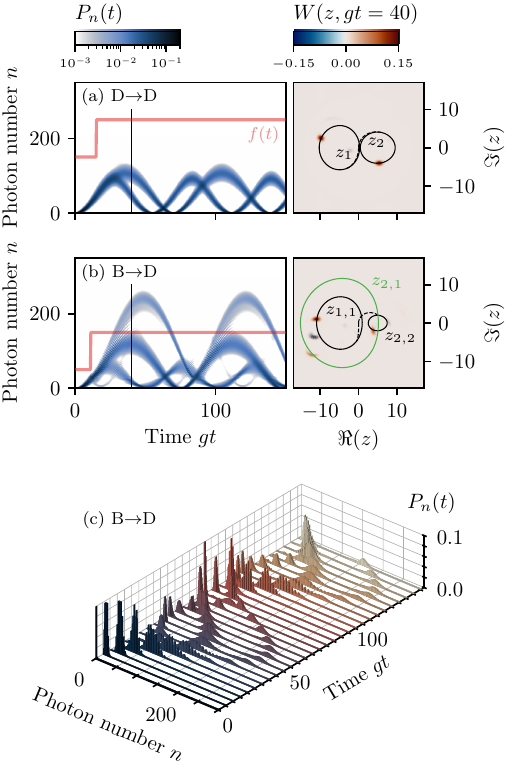}
	\caption{
		Photon number distribution $P_{n}(t)$ resolved in photon number $n$ as well as time $t$ (left) with a single step in the driving strength using the parameters (a) $f_{0} = 15\, g$, $f_{1} = 25\, g$, $g \tau = 15$ and (b) $f_{0} = 5\, g$, $f_{1} = 15\, g$, $g \tau = 11$.
		In both cases a detuning of $\delta = 0.1\, g$ was chosen.
		The red line sketches $f(t)$ while the black line marks the time $g t = 40$, at which the Wigner function $W(z)$ is shown (right).
		The dynamics in phase space of the initial packets $z_{1}$, $z_{2}$ and their continuations after the step $z_{1,1}$, $z_{2,2}$, obtained through Eq.~\eqref{eqn:variational_eom_z_adiab} as described in Sec.~\ref{ssec:single}, are given shown as black lines in the right panels.
		The trajectory $z_{2,1}$ of the packet generated by increasing the driving strength is given by the green line.
		An animation of the time evolution of $W(z, t)$ can be found in the supplementary material.
		(c) The same data as in (b) in a three-dimensional view.
	}
	\label{fig:single}
\end{figure}

Our ultimate goal is to find a reliable scheme to generate a controlled number of packets in the photon statistics.
Therefore, we exclude class A from our considerations, since in this regime both packets experience significant dispersion [cf. Fig.~\ref{fig:constant_f}(a)].
When we restrict $f_{0}$ and $f_{1}$ to classes B, C and D, $\varphi_{1}$ consists of only $\Pp{}$ to good approximation irrespective of the driving strength.
As a result, $\Psi_{1}$ remains largely unchanged by the change in driving strength.
The same holds true for $\Psi_{2}$ if both $f_{0}$ and $f_{1}$ fall into class D and hence the overall packet structure remains unchanged in this case, i.e. we have two packets all the time and switching the driving strength did not generate an additional packet [cf. Fig.~\ref{fig:single}(a)].

The simplest situations to analyze are B$\rightarrow$D steps, i.e. choosing $f_{0}$ within B and $f_{1}$ within D.
In these cases, any contribution of $\Pp{}$ to $\Psi_{2}(\tau)$ forms a third packet evolving according to the laws of $\varphi_{1}$ while the remaining part, constructed of $\Pm{}$, continues its evolution according to the laws of $\varphi_{2}$.
An example of this is shown in Fig.~\ref{fig:single}(b).
The photon number distribution $P_{n}(t)$ confirms our expectation of it largely consisting of three distinct packets.
The interpretation in terms of the simplified variational theory is verified by inspecting the Wigner function.
Disregarding the parts of rapidly oscillating sign indicating coherences, $W$ is composed of three isolated parts (for $t > \tau$) corresponding to the three packets in $P_{n}$.
The trajectories of these parts can be reproduced by solutions of Eq.~\eqref{eqn:variational_eom_z_adiab} in the following way.
We calculate $z_{1}(t)$ and $z_{2}(t)$ by numerical integration with driving strength $f_{0}$ for $0 < t < \tau$ with initial states $z_{1/2}(0) = 0$.
Then, $z_{j,k}(t)$ for $t > \tau$ is obtained by integrating the equation corresponding to $\varphi_{k}$ with driving strength $f_{1}$ and initial state $z_{j}(\tau)$.

In order to better see the evolution of the shapes of the wave packets, we again show the data from Fig.~\ref{fig:single}(b) in a three-dimensional plot in Fig.~\ref{fig:single}(c).
As can be seen from this figure, for the chosen parameters the packets do not disperse much over extended period of time.

\subsection{Multiple steps}
\label{ssec:multiple}
Given the success in generating a single additional packet, it is natural to ask if the process can be repeated.
We introduce even more steps in the driving function.
Formally, we can write
\begin{align}
	f(t) = \sum_{j} f_{j}\, (\Theta(t-\tau_{j}) - \Theta(t-\tau_{j+1}))
\end{align}
where $\Theta(t)$ is the Heaviside step function.
We focus on the most promising situations, namely those in which $f_{0}$ is chosen in accordance with class B and $f_{1}$ in D.
Further increasing the driving strength results in $f_{2}$ corresponding to class D as well.
Thus, the packet structure is not influenced by the second step as was described in the previous section.
We investigate two strategies to induce several packets circumventing this challenge.

\begin{figure}[t]
	\centering
	\includegraphics[scale=1]{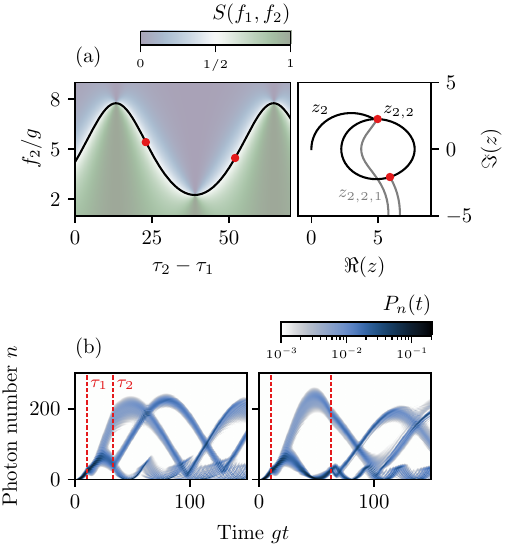}
	\caption{
		(a) Variational approximation with two steps in the driving function obtained by numerically integrating Eq.~\eqref{eqn:variational_eom_z_adiab} using $f_{0} = 5\, g$, $g \tau_{1} = 10.5$ and several values of $f_{2}$ and $\tau_{2}$.
		The left panel shows the overlap $S(f_{1}, f_{2})$ [cf. Eq.~\eqref{eqn:variational_overlap}] as a function of $f_{2}$ and $\tau_{2}$ evaluated at $z_{2, 2}(\tau_{2})$, i.e. the trajectory of the remaining packet in state $\varphi_{2}$ after the step at $\tau_{1}$.
		The black line indicates the pairs $(\tau_{2}, f_{2})$ for which $S(f_{1}, f_{2}) = 1/2$.
		The two examples $(g (\tau_{2}-\tau_{1}) = 23, f_{2} = 5.4\, g)$ and $(g (\tau_{2}-\tau_{1}) = 52, f_{2} = 4.5\, g)$ are marked by the red dots.
		The right panel plots the trajectory of the remaining $\varphi_{2}$-packet as well as those of the generated packet after the step at $\tau_{2}$ (gray lines) for the two examples mentioned above.
		(b) Photon number distribution $P_{n}(t)$ resolved in photon number $n$ as well as time $t$ obtained by solutions of Eq.~\eqref{eqn:schrodinger} using the initial state $\Psi(0) = \Pm{0}$ and the parameters of the two examples in (a).
		The dashed red lines indicate the points in time of the steps $\tau_{1}$ and $\tau_{2}$.
	}
	\label{fig:multiple_1}
\end{figure}

\begin{figure}[t]
	\centering
	\includegraphics[scale=1]{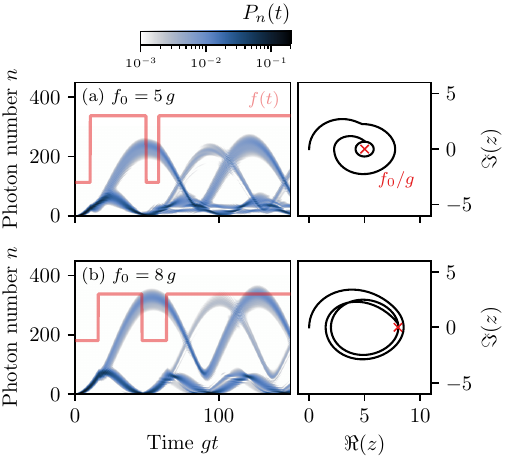}
	\caption{
		Photon number distribution $P_{n}(t)$ resolved in photon number $n$ as well as time $t$ (left) with several steps in the driving strength.
		The results were obtained by numerically integrating Eq.~\eqref{eqn:schrodinger} with the initial state $\Psi(0) = \Pm{0}$ using the parameters (a) $f_{0} = f_{2} = 5\, g$, $f_{1} = f_{3} = 15\, g$, $g \tau_{1} = 10.5$, $g \tau_{2} = 49.4$, $g \tau_{3} = 58.2$ and (b) $f_{0} = f_{2} = 8\, g$, $f_{1} = f_{3} = 15\, g$, $g \tau_{1} = 16.1$, $g \tau_{2} = 46.5$, $g \tau_{3} = 63.7$.
		In both cases a detuning of $\delta = 0.1\, g$ is chosen.
		The red line sketches the time dependent driving strength $f(t)$.
		The trajectory of the remaining packet in state $\varphi_{2}$  within the variational approximation (right) was calculated by numerically integrating Eq.~\eqref{eqn:variational_eom_z_adiab} using the same parameters.
		The red cross marks the point $z = f_{0} / g$.
	}
	\label{fig:multiple_2}
\end{figure}

First, we follow the most straightforward plan of directly using a reduction in driving strength to generate another packet.
Suppose our goal was to split $\Psi_{2}$ into three packets of equal contribution.
We can use the variational approximation to easily determine the values for $\tau_{1}$, $\tau_{2}$ and $f_{2}$ necessary to do so.
With fixed choices of $f_{0}$ and $f_{1}$ and the initial state $z_{2}(0) = 0$, we solve Eq.~\eqref{eqn:variational_eom_z_adiab} until a point in time is reached at which the overlap
\begin{align}\label{eqn:variational_overlap}
	S(f_{0}, f_{1})
	= |\langle \varphi_{2}(f_{0}),\, \varphi_{1}(f_{1})\rangle|^{2}
\end{align}
evaluated at $z_{2}(t)$ is equal to $1/3$.
This time fixes $\tau_{1}$ splitting $\Psi_{2}$ into $\Psi_{2,1}$ and $\Psi_{2,2}$ with relative weights of $1/3$ and $2/3$.
Next, we again solve Eq.~\eqref{eqn:variational_eom_z_adiab} using the driving strength $f_{1}$ and initial state $z_{2,2}(0) = z_{2}(\tau_{1})$ and observe at which time $S(f_{1}, f_{2})$ evaluated at $z_{2,2}(t)$ takes the value $1/2$.
Choosing $\tau_{2}$ at precisely this point in time splits $\Psi_{2,2}$ into two packets $\Psi_{2,2,1}$ and $\Psi_{2,2,2}$ of equal contribution.
The optimal choice for $\tau_{2}$ in general depends on the value of $f_{2}$, as is visualized in Fig.~\ref{fig:multiple_1}(a).
Since there are no additional restrictions on $f_{2}$, we can choose it freely in order to establish a desired delay $\tau_{2} - \tau_{1}$ between the two generated packets.
However, the freedom in $\tau_{2}$ is somewhat restrained by the observation, that we ought not to choose the proposed optimal values if $z_{2,2}(\tau_{2})$ is close to the upper or lower turning point of its oscillation.
In this case, $z_{2,2}(\tau_{2}) \approx f_{2}/g$ and thus the packets would traverse this point leading to repeated splits analogous to the behavior in class C.
Numerical calculations, shown in Fig.~\ref{fig:multiple_1}(b), with parameters chosen precisely in the described way, validate the methodology.

The second strategy is motivated by the desire to maximize the time spent in class D, which stems from the fact that the packet dispersion is smaller if the LDS are decoupled \cite{nimmesgern_multiple_2024}.
The basic idea is to generate all further packets by B$\rightarrow$D steps, such that propagation after the step continues in class D.
We have to include a D$\rightarrow$B step beforehand, which is chosen such that the packet structure is unaffected at this step, i.e. when the remaining packet in state $\varphi_{2}$ reaches its lower turning point.
Note that in contrast to the previous strategy, we lack control of the delay between the packets without affecting their relative weights.
In Fig.~\ref{fig:multiple_2}, we show two examples using this method for different values of $f_{0} = f_{2}$.
The higher $f_{0}$, the less the remaining $\varphi_{2}$-packet itself and its trajectory in phase space is deformed.
However, there is a drawback in that the point $z = f_{0} / g$ in phase space is closer to the outside of the trajectory.
This limits the extensibility of this strategy to generate an even larger number of packets, since if the trajectory fails to enclose this point, the overlap $S(f_{0}, f_{1})$ is small at all times (cf. Sec.~\ref{ssec:single}).

The protocols presented above still leave open potential possibilities for optimization.
Further, the system allows for many more different strategies to directly affect its state.
Too many, in fact, to give a comprehensive account in this work.
However, the general methodology described in the above two examples should make the task easy to get a lucid picture of the system's response to any chosen protocol.

\section{Extraction of the packet structure from the mean photon number}
\label{sec:measurement}

\begin{figure}
	\centering
	\includegraphics[scale=1]{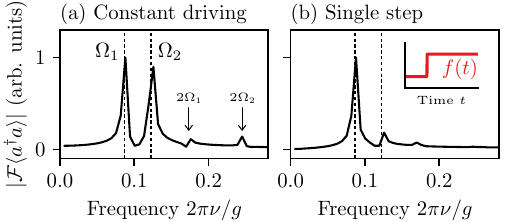}
	\caption{
		Discrete Fourier transform of the mean photon number $\mathcal{F} \langle a^{\dagger} a \rangle$ obtained via numerical solutions of Eq.~\eqref{eqn:schrodinger} for a simulation time of $g T = 1000$.
		The detuning was chosen as $\delta = 0.1\, g$ and the driving strength as (a) constant $f = 15\, g$ and (b) with a single step, as illustrated by the red line, using $f_{0} = 5\, g$, $f_{1} = 15\, g$ and $g \tau = 11$.
		The dashed lines indicate the oscillation frequencies $\Omega_{1/2}$ in class D as given by Eq.~\eqref{eqn:variational_freq}.
	}
	\label{fig:measurement}
\end{figure}

Even though modern experimental techniques allow the full photon number distribution to be measured~\cite{schlottmann_exploring_2018, schmidt_photon-number-resolving_2018, helversen_quantum_2019}, in particular its time-resolved measurement remains challenging.
However, the composition of $P_{n}$ in terms of photon number wave packets can be determined via the dynamical evolution of the mean photon number $\langle a^{\dagger} a \rangle$, which is much easier to obtain.
The different incommensurable frequencies at which the individual packets travel are inherited to $\langle a^{\dagger} a \rangle$.
Thus, their presence can be identified in the peaks of the Fourier transform $\mathcal{F} \langle a^{\dagger} a \rangle$, which in general are distinguishable from the harmonics of a single fundamental frequency [cf. Fig.~\ref{fig:measurement}(a)].
For these peaks to be well-defined, the packets have to perform at least a couple of oscillation cycles.
Class D is hence best-suited for this purpose due to the lack of significant packet dispersion.
In this regime, the oscillation frequencies are given by (cf. Appendix~\ref{app:frequency})
\begin{align}\label{eqn:variational_freq}
	\Omega_{1/2}
	= \frac{\delta}{\sqrt{1 \pm g^{2} / 2 f \delta}}.
\end{align}

The additional packets generated by steps in the driving strength correspond to $\varphi_{1}$ and thus fall in the LDS decoupled regime in classes B, C and D.
Here, the oscillations are harmonic in nature, i.e. the frequency is independent of the amplitude (cf. Appendix~\ref{app:solution}).
Therefore, these packets $\Psi_{2, 1}$, $\Psi_{2, 2, 1}$, etc. oscillate with the same frequency as $\Psi_{1}$ and, as a consequence, do not constitute new peaks in $\mathcal{F} \langle a^{\dagger} a \rangle$.
However, their generation can be identified by a relative increase of the peak at $\Omega_{1}$ with respect to that at $\Omega_{2}$, as is shown in Fig.~\ref{fig:measurement}(b).

\section{Summary}
\label{sec:summary}
In this paper we investigated the possibility to manipulate the number of photon number wave packets in a single-mode cavity.
We first revisited the generation of photon number wave packets by resonant constant driving of a two-level system coupled to the cavity.
Using an approximation based on a time-dependent variational principle, we were able to reduce the number of relevant degrees of freedom to two, while still retaining a quantitatively accurate description.
Our major new insight is the identification of four different dynamical regimes which are distinguished by different compositions of the TLS parts associated with the photon number wave packets.
The different regimes can be selected by suitable choices of the driving strength and the detuning.

We then extended the discussion to driving the system by a laser where the driving strength exhibits one or more sudden switches.
It turns out that a sudden switch may or may not change the number of simultaneous photon number wave packets.
We found out that the mechanism for packet generation depends crucially on the TLS states of the individual packets.
Detailed analysis of the pertinent conditions can be simplified by referencing the dynamical regimes of constant driving.
Most importantly, based on our analysis we were able to develop protocols that allow for repeated on demand generation of additional photon number wave packets.

Finally, we discussed the challenge of ascertaining the packet structure of the photon number distribution.
While time-resolved measurements of the full photon number statistics remain a challenging task, the mean photon number is easy to obtain.
We identified a method to extract the qualitative packet structure merely from the time-dependence of the mean photon number.

The methods developed in this work provide a new toolkit to shape the structure of the photon number distribution at will.
Such in principle simple methods suggest a promising perspective for new ways of quantum information processing.
The analytical methods used to investigate the proposed protocols constitute a framework to analyze different and more complicated protocols.
Furthermore, it is possible to extend these methods to analyze for instance effects of dissipation.

\appendix

\section{Derivation of the variational approximation}
\label{app:variational}
Given an Ansatz for the time-dependent wave function that is described by a set of parameters, we can obtain an optimal approximation to the solution of the Schr\"{o}dinger equation on the subspace spanned by the Ansatz functions by extremizing the functional
\begin{align}
	I = \int \mathrm{d}t\ \langle \Psi | i \hbar\, \dot{\Psi} - H \Psi \rangle
\end{align}
with respect to the parameters~\cite{broeckhove_equivalence_1988}.
Using the Hamiltonian Eq.~\eqref{eqn:hamiltonian} and the Ansatz Eq.~\eqref{eqn:variational_ansatz}, we find
\begin{multline}
	I
	= \int \mathrm{d}t\ \Bigl(i \hbar\, \alpha^{*} \dot{\alpha} + i \hbar\, \beta^{*} \dot{\beta}
	+ \frac{i \hbar}{2}\, (z^{*} \dot{z} - z \dot{z}^{*})
	\\
	- \hbar \delta\, |z|^{2} - \hbar g\, \bigl(z \alpha \beta^{*} + z^{*} \alpha^{*} \beta\bigr)
	\\
	+ \hbar f\, \bigl(\alpha \beta^{*} + \alpha^{*} \beta\bigr)\Bigr).
\end{multline}
Varying this expression with respect to $\alpha^{*}$, $\beta^{*}$ and $z^{*}$ leads to the equations of motion
\begin{subequations}
\begin{align}
	0 &= i \hbar\, \dot{\alpha} - \hbar g\, z^{*} \beta + \hbar f\, \beta
	\\
	0 &= i \hbar\, \dot{\beta} - \hbar g\, z \alpha + \hbar f\, \alpha
	\\
	0 &= i \hbar\, \dot{z} - \hbar \delta\, z - \hbar g\, \alpha^{*} \beta,
\end{align}
\end{subequations}
which coincide with Eqs.~\eqref{eqn:variational_eom}.

As discussed in Sec.~\ref{sec:constant_f}, the solution of the variational equations of motion can be well approximated by the instantaneous eigenstates of the TLS equations.
These are obtained by diagonalizing $H_\mathrm{TLS}(z)$, as defined in Eq.~\eqref{eqn:variational_HTLS}, treating $z$ as constant:
\begin{align}
	H_{\mathrm{TLS}}(z) \varphi_{1/2} = \hbar \omega_{1/2}\, \varphi_{1/2}.
\end{align}
The eigenfrequencies are easily obtained as given in Eq.~\eqref{eqn:variational_eigenfrequency}.
Using this expression, the determining equation of the ground and excited state amplitudes $\varphi_{1/2}^{\mathrm{G}}$ and $\varphi_{1/2}^{\mathrm{X}}$ of the eigenvectors read
\begin{align}\label{eqn:app:eigenvector}
	(g z - f)\, \varphi_{1/2}^{\mathrm{G}} = \mp |g z - f|\, \varphi_{1/2}^{\mathrm{X}}.
\end{align}
Taking the absolute value of this equation, we are immediately lead to the result $|\varphi_{1/2}^{\mathrm{G}}| = |\varphi_{1/2}^{\mathrm{X}}|$, which together with the normalization condition leads to
\begin{align}
	|\varphi_{1/2}^{\mathrm{G}}| = |\varphi_{1/2}^{\mathrm{X}}| = \frac{1}{\sqrt{2}}.
\end{align}
To determine the relative phase of the amplitudes, we multiply Eq.~\eqref{eqn:app:eigenvector} by $(\varphi_{1/2}^{\mathrm{G}})^{*}$ resulting in
\begin{align}\label{eqn:app:eigenvector_phase}
	(\varphi_{1/2}^{\mathrm{G}})^{*} \varphi_{1/2}^{\mathrm{X}}
	&= \mp \frac{1}{2} \frac{g z - f}{|g z - f|}
\end{align}
The adiabatic solutions of $(\alpha, \beta)$ differ from the eigenvectors $\varphi_{1/2}$ only by a multiplicative complex phase [cf. Eq.~\eqref{eqn:variational_adiabatic}].
Thus, within the adiabatic approximation $\alpha^{*} \beta = (\varphi_{1/2}^{\mathrm{G}})^{*} \varphi_{1/2}^{\mathrm{X}}$ and inserting this expression into the equation of motion for $z$ directly results in Eq.~\eqref{eqn:variational_eom_z_adiab}.
On top of that, we can find $\lambda_{1/2}$ by evaluating the real part of Eq.~\eqref{eqn:app:eigenvector_phase} as follows
\begin{align}
	\lambda_{1/2}(z)
	&= 2 \Re((\varphi_{1/2}^{\mathrm{G}})^{*} \varphi_{1/2}^{\mathrm{X}})
	\nn
	&= \mp \mathrm{sgn}(\Re(z) - f/g) \Bigl(1 + \frac{\Im(z)^{2}}{(\Re(z) - f/g)^{2}}\Bigr)^{-1/2}
\end{align}
producing Eq.~\eqref{eqn:variational_lds}.

\section{Turning points of $z(t)$ within the adiabatic approximation}
\label{app:turning}
The solutions of Eq.~\eqref{eqn:variational_eom_z_adiab} for $\delta \neq 0$ cannot be expressed analytically in simple fashion.
However, we are able to obtain key features of the trajectories.
To this end, we make continued use of conservation of energy which, using Eqs.~\eqref{eqn:variational_Hconst} and \eqref{eqn:variational_Hvalue}, results in the condition
\begin{align}\label{eqn:app:const}
	0 = \hbar \delta\, |z_{1/2}|^{2} \mp \hbar\, |g z_{1/2} - f| \pm \hbar f.
\end{align}

Eq.~\eqref{eqn:variational_eom_z_adiab} is symmetric under reflection along the real axis $\Im(z) \rightarrow -\Im(z)$ combined with time reversal $t \rightarrow -t$.
As a result, the solution trajectories are mirror symmetric with respect to the real axis.
This culminates in the fact that the turning points $\tilde{z}_{1/2}$ are purely real.
The same can be established by inspecting the equation of motion for the mean photon number $|z|^{2}$:
\begin{align}
	\frac{\mathrm{d}}{\mathrm{d}t} |z_{1/2}|^{2} = \mp f \frac{\Im(z)}{|z - f/g|}.
\end{align}

In order to find expressions for the turning points, we evaluate Eq.~\eqref{eqn:app:const} for real $z_{1/2}$.
Assuming $z_{1/2} < f/g$, it takes the form
\begin{subequations}
\begin{align}
	0 = \delta\, z_{1/2}^{2} \pm g\, z_{1/2}
\end{align}
\end{subequations}
with solutions $z_{1/2} = 0$ and $z_{1/2} = \mp g / \delta$.
Note that the latter solution for $z_{2}$ only fulfills the starting assumption $z_{2} < f/g$ if $\delta < g^{2} / f$ whereas the former solution as well as both solutions for $z_{1}$ are consistent for any parameters (if $\delta > 0$).
If, on the other hand, $z_{1/2} > f/g$, we have
\begin{align}
	0 = \delta\, z_{1/2}^{2} \mp g\, z_{1/2} \pm 2 f,
\end{align}
admitting the solutions
\begin{subequations}
\begin{align}
	z_{1} &= \frac{g}{2 \delta} \Biggl(\pm \sqrt{1 - \frac{8 f \delta}{g^{2}}} + 1\Biggr),
	\\
	\label{eqn:z2_roots}
	z_{2} &= \frac{g}{2 \delta} \Biggl(\pm \sqrt{1 + \frac{8 f \delta}{g^{2}}} - 1\Biggr).
\end{align}
\end{subequations}
These expressions for $z_{1}$ are only real if $\delta \leq g^{2} / 8 f$.
The first solution for $z_{1}$ decreases monotonically with increasing $\delta$ reaching $4 f / g$ at $\delta = g^{2} / 8 f$, whereas the second solution increases monotonically from $2 f / g$ to $4 f / g$.
Therefore both solutions are consistent with the foregoing assumptions and hence Eq.~\eqref{eqn:app:const} has four real solutions in the parameter range $0 < \delta \leq g^{2} / 8 f$.
Since Eq.~\eqref{eqn:variational_eom_z_adiab} is a first order differential equation, the trajectories of its solutions cannot intersect.
As a result, the inner two and the outer two real solutions of Eq.~\eqref{eqn:app:const} respectively are connected on a single trajectory.
In summary,
\begin{align}\label{eqn:app:turning_1}
	\tilde{z}_{1} = \begin{cases}
		\displaystyle \frac{g}{2 \delta} \Biggl(1 - \sqrt{1 - \frac{8 f \delta}{g^{2}}}\Biggr)
		& \text{if } \delta \leq g^{2} / 8 f
		\\[5mm]
		\displaystyle -\frac{g}{\delta}
		& \text{if } \delta > g^{2} / 8 f
	\end{cases}.
\end{align}
The situation is simpler in the case of $z_{2}$ as its second solution is never greater than $f / g$ whereas the first solution fulfills this condition precisely if $\delta < g^{2} / f$.
Hence,
\begin{align}\label{eqn:app:turning_2}
	\tilde{z}_{2} = \begin{cases}
		\displaystyle \frac{g}{2 \delta} \Biggl(\sqrt{1 + \frac{8 f \delta}{g^{2}}} - 1\Biggr)
		& \text{if } \delta \leq g^{2} / f
		\\[5mm]
		\displaystyle \frac{g}{\delta}
		& \text{if } \delta > g^{2} / f
	\end{cases}.
\end{align}

\section{Parameter regimes of the dynamical classes}
\label{app:classes}
The classes A, B and D are defined by one or both packets being approximately restricted to a subspace spanned by a single LDS (cf. Sec.~\ref{sec:constant_f}).
In order to obtain boundaries in parameter space between these classes, we derive expressions for the extrema of $\lambda$ on the trajectory $z(t)$.
Using Eqs.~\eqref{eqn:app:eigenvector_phase} and \eqref{eqn:app:const}, we can can write $\lambda$ as a function of $|z - f/g|$ on the photonic trajectory:
\begin{multline}\label{eqn:app:lds}
	\lambda_{1/2}
	= \mp\frac{\Re(z_{1/2}) - f/g}{|z_{1/2} - f/g|}
	\\
	= \pm\frac{g}{2 f} |z_{1/2} - f/g| - \frac{g^{2}}{2 f \delta}
	\\
	\pm \frac{g}{2 \delta} \Bigl(\frac{f \delta}{g^{2}} \pm 1\Bigr) \frac{1}{|z_{1/2} - f/g|}.
\end{multline}

For $\delta > g^{2}/8f$, $\lambda_{1}$ is greater than $-1$ at all times.
Minimizing the expression in Eq.~\eqref{eqn:app:lds}, we find in this range
\begin{align}
	\min \lambda_{1} = \sqrt{1 + \frac{g^{2}}{f \delta}} - \frac{g^{2}}{2 f \delta}.
\end{align}
Inverting this relation, we find that in order to ensure that $\min \lambda_{1} > 1 - \eta$ for some desired $\eta$, we have to choose
\begin{align}
	\frac{f \delta}{g^{2}} > \frac{1}{2 \eta + \sqrt{8 \eta}}.
\end{align}
A precise choice for $\eta$ is somewhat arbitrary in accordance with the transitions between classes being continuous rather than well-defined at a specific set of parameters.
For the sake of definiteness, we take $\eta = 1/2$, such that $f \delta = g^{2} / 3$ is determined as the boundary between classes A and B.
Note that on this boundary $\lambda_{1} = 1/2$ is decidedly its minimal value.
The average is significantly closer to $1$, meaning that the dominate constituent of the states is the upper LDS.

Similarly, for $\delta > g^{2}/f$, $\lambda_{2}$ is smaller than $1$ at all times.
According to Eq.~\eqref{eqn:app:lds} its maximum is given by
\begin{align}
	-\max \lambda_{2} = \sqrt{1 - \frac{g^{2}}{f \delta}} + \frac{g^{2}}{2 f \delta}.
\end{align}
Thus, all parameters with $f \delta > g^{2}$ satisfy $\max \lambda_{2} < -1/2$ and can be classified as belonging to class D insofar the adiabatic approximation is valid.
However, it is known that for $f \delta \approx g^{2}$ the trajectory $z_{2}(t)$ approaches the point $f/g$ at its oscillation maximum [cf. Eq.~\eqref{eqn:app:turning_2}], where the adiabatic approximation fails.
Therefore, class C lies between and represents the boundary of classes B and D.

\section{Solution of Eq.~\eqref{eqn:variational_eom_z_adiab} in two limiting cases}
\label{app:solution}
In the case of vanishing detuning, Eq.~\eqref{eqn:variational_eom_z_adiab} can be readily integrated by expressing $z - f/g$ in terms of its magnitude and phase
\begin{align}
	z(t) - f/g = r(t) e^{i \vartheta(t)}
\end{align}
Then, the equation of motion takes the form
\begin{align}
	i \hbar\, \dot{r}_{1/2} - \hbar\, r_{1/2} \dot{\vartheta}_{1/2}
	&= \mp \frac{\hbar g}{2}
\end{align}
which is immediately solved by
\begin{subequations}\label{eqn:app:nodet_sol}
\begin{align}
	r_{1/2}(t) &= R
	\\
	\vartheta_{1/2}(t) &= \Theta \pm \frac{g}{2 R}\, t
\end{align}
\end{subequations}
with constants $R$ and $\Theta$.
With the initial condition $z(0) = 0$, we thus obtain
\begin{align}
	z_{1/2}(t) = \frac{f}{g} \Bigl(1 - \exp\Bigl(\pm i\, \frac{g^{2}}{2 f}\, t\Bigr)\Bigr)
\end{align}
consistent with the results of Ref.~\cite{chough_nonlinear_1996}.

In the opposite limiting case of large detuning, $\lambda_{1}$ ($\lambda_{2}$) remains close to $-1$ ($1$) throughout the evolution.
Thus, we can expect the imaginary part of $z - f/g$ to be much smaller than its real part [cf. Eq.~\eqref{eqn:variational_lds}] and we can approximate $(z - f/g) / |z - f/g|$ by $-1$ (since $z < f/g$).
Then, Eq.~\eqref{eqn:variational_eom_z_adiab} takes the form
\begin{align}
	i \hbar\, \dot{z}_{1/2} = \hbar \delta\, z_{1/2} \pm \frac{\hbar g}{2}
\end{align}
with solutions
\begin{align}\label{eqn:app:highdet_sol}
	z_{1/2}(t) = \mp \frac{g}{2 \delta} \Bigl(1 - \exp(-i \delta t)\Bigr)
\end{align}
consistent with the approximations in the LDS-decoupled regime shown in Ref.~\cite{nimmesgern_multiple_2024}.
In contrast to the anharmonic oscillations for $\delta = 0$, the oscillations for large detuning are harmonic in nature as the frequency no longer depends on the amplitude [cf. Eqs.~\eqref{eqn:app:nodet_sol}].

\section{Oscillation frequency in class D}
\label{app:frequency}
Eq.~\eqref{eqn:app:highdet_sol} is the solution of Eq.~\eqref{eqn:variational_eom_z_adiab} in the limiting case $\delta \rightarrow \infty$.
In this limit, both packets exhibit the same oscillation frequency
\begin{align}
	\Omega_{1/2} \sim \delta.
\end{align}
However, in general $\Omega_{1}$ and $\Omega_{2}$ do not coincide in class D as numerical solutions indicate [cf. Fig.~\ref{fig:constant_f}(a)].
Corrections to the frequency can be obtained systematically by expanding $T = \pi / \Omega$, which we define as half the oscillation period, in orders of $1/\delta$.
Formally, $T$ can be expressed as
\begin{align}\label{eqn:app:T_integral}
	T
	= \int_{\nu(0)}^{\nu(T)} \frac{\mathrm{d}\nu}{\dot{\nu}},
\end{align}
where $\nu = |z - f/g|^{2}$.
With the initial condition $z(0) = 0$, we have $\nu(0) = (f/g)^{2}$ and Eqs.~\eqref{eqn:app:turning_1} and \eqref{eqn:app:turning_2} yield the upper limit of integration $\nu_{1/2}(T) = (f/g \pm g/\delta)^{2}$.
The time derivative of $\nu$ is given by
\begin{align}
	\dot{\nu} = -\frac{2 f \delta}{g}\, \Im(z).
\end{align}
The final ingredient necessary in the integral is an expression of $\Im(z)$ in terms of $\nu$, which is obtained via Eq.~\eqref{eqn:app:const},
\begin{align}
	\Im(z_{1/2})
	&= \mp\sqrt{\nu - \frac{g^{2}}{4 f^{2}}\Bigl(\nu + \frac{f^{2}}{g^{2}}
	\mp \frac{g}{\delta} \sqrt{\nu} \pm \frac{f}{\delta}\Bigr)^{2}}.
\end{align}
By using the substitution
\begin{align}
	\nu = \frac{f^{2}}{g^{2}} + x \Bigl(\frac{g^{2}}{\delta^{2}} \pm \frac{2 f}{\delta}\Bigr)
\end{align}
and subsequently expanding the integrand in orders of $1/\delta$, Eq.~\eqref{eqn:app:T_integral} can be shown to give
\begin{multline}
	T_{1/2}
	= \frac{1}{\delta} \sqrt{1 \pm \frac{g^{2}}{2 f \delta}}
	\\
	\times \int_{0}^{1} \mathrm{d}x \Bigl((x - x^{2}) + \mathscr{O}(\delta^{-3})\Bigr)^{-1/2}.
\end{multline}
The remaining integral in $x$ is equal to $\pi$, such that we finally obtain
\begin{multline}
	\Omega_{1/2}
	= \frac{\delta}{\sqrt{1 \pm g^{2} / 2 f \delta}} + \mathscr{O}(\delta^{-2})
	\\
	= \delta \mp \frac{g^{2}}{4 f} + \frac{3}{32} \frac{g^{4}}{f^{2} \delta} + \mathscr{O}(\delta^{-2}).
\end{multline}

\bibliography{bib}

\end{document}